\documentclass[aps,preprint,showpacs,prl]{revtex4}
\usepackage{graphics}

\begin{document}

\title{Now You See It, Now You Don't---The Pattern
of Production of Certain Resonances }

\author{Shmuel Nussinov$^{(a)}$}
\email{nussinov@post.tau.ac.il} \affiliation{Department of
Physics \& Astronomy,
University of South Carolina, Columbia, USA\\
$^{(a)}$on sabbatical leave from School of Physics and Astronomy,
Tel Aviv University, Ramat-Aviv, Tel Aviv 69978, Israel}

\date{August 5, 2004}

\begin{abstract}
  We try to motivate from  QCD a pattern of production in various reactions
 of (non)exotic resonances. A higher penalty for extra $q\bar{q}$ production
 in $e^+ e^-$ collisions than in collisions with a nucleon target may explain
 the absence of exotic multi-quark states in $e^+ e^-$ .
 We also briefly address the doubly charmed baryons and the utilization
 of QCD inequalities in connection with possible new hadronic states.
\end{abstract}

\maketitle

 \section{Introduction}

  The issue of selective production of various hadronic states under different circumstances
 has renewed and enhanced urgency. It is particularly the case for the 1540 MeV pentaquark.
 By now the $K^+n$ or $K^0p$ resonance has been seen in many $\gamma$ proton  and neutrino proton
 (or deutron/nuclei)experiments\cite{Seen}. It has however not been seen in
 several $e^+e^-$ and
 $p-\bar{p}$ collider experiments to be reported in the upcoming ICHEP.
 A related example is that of the new $D_s(2362)$ seen by the Selex hyperon beam
 charm experiment
 in both $D_s \eta$ and $D^0 K^+$ decay modes but not found in $e^+ e^-S$  colliders.
 Selex found also evidence for doubly charmed baryons which were not confirmed by
 other experiments.

 In many cases in the past some scientists discovered something new, which others, using the same or
 different methodology, failed to verify. Every conflict was eventually settled and a consensus reached.
 In most cases the conservatives prevailed and the apparent new discovery was discarded.
 In a handful of cases {\it all} experiments were correct and a novel idea resolved the apparent conflicts.
 A prime, recent example is that of the Solar neutrinos. The Vacuum {\it and} MSW effects combine to account for
 the perplexing pattern of solar neutrinos observed in different experiments with different sensitivity for
 neutrinos of different energies and flavors.\cite{Ba}

  We wish to point out that for the exotic pentaquark and a ``crypto-exotic"
 $D_s (2632)$ the second possibility may not be ruled out.

\section{The Argument for Different Production Rates  of Exotic and Non-Exotic
Hadrons in $e^+ e^-$ Collisions}

  The BaBar and Belle $e^+ e^-$ colliders at $ W=m (\Upsilon \, 4S) \sim$ 10.5  GeV, were designed
 to study the physics of heavy (b) flavor decays and  CP violation. Yet these experiments may be
 the best tool for studying hadrons made of uds and c quarks and mesons in particular.

 Jointly BaBar and Belle will collect several billion  $e^+ e^- \rightarrow c\bar{c}, s\bar{s},
 d\bar{d}$ and $u\bar{u}$ collisions. With $4\pi$ geometry, excellent momentum and angular resolution
 and good particle identification these experiment have reconstructed all known (mesonic) resonances
 and a few more, and so far this has mainly been a by-product of studying $B$ decays!

  The following pattern is theoretically expected for $ e^+ e^- $ colliders:
 Ordinary non-exotic resonances made of a quark and an anti-quark are more copious than exotic resonances made
 of two quarks and two anti-quarks. This can help determine the quark content of the state under study.

  Our claim may seem surprising. On average $\sim$ ten pions are produced in $e^+ e^-$ collisions
 10.5 GeV. Viewing each pion as a $q\bar{q}$ pair we have ten light quarks, and ten antiquarks produced
 in each collision. In this case there is no preference of $q\bar{q}$ non-exotics with two quarks
 and antiquarks relative to $q\bar{q} \, q\bar{q}$ exotics with altogether four quarks and anti-quarks.

 This simplistic argument is wrong and the preferred production  of non-exotics in electron
 positron collisions is model independent, tracing back to the $1/N_c$ expansion in QCD.

  Thus let us follow the evolution of the system. After the $e^+ e^- \rightarrow$
 virtual $\gamma \rightarrow c\bar{c} $ or $q{\bar{q}}$ with  $q=u,d,s$ collision,
  the primary high energy quark and anti-quark emerge as highly virtual ``hot lines".
 Subsequent evolution ``splits" off additional gluons and quark-anti-quark pairs with
 decreasing virtuality and energy which is shared by an increasing number of constituents.
 Eventually confinement sets in and rather than use the picture of quark and
 gluons we revert to hadrons: $q\bar{q}$ non-exotic mesons, $qqq$ baryons
 (and antibaryons), glueballs and  potentially also $q\bar{q}q\bar{q}$ exotics.

   At what stage does this happen and what is the average mass $M$ of these primordial hadronic
 clusters? Several arguments suggest that $M \sim$ 1.2-2 GeV \cite{Weinstein}. The perturbative
 evolution proceeds until a sufficient number of extra, light quark pairs  have been produced so
 that the invariant masses of  $q_i\bar{q}_j$ which are neighboring in rapidity is $\sim M$.
 These clusters are {\it color singlets} made of quarks and anti-quarks or glueballs. The mass of
 the lightest glueballs $\sim$ 1.5-2 GeV confirms the above estimate of $M$. The glueballs and the
 highly excited mesons are broad and quickly decay into lighter $q\bar{q}$ mesons and pions in particular.

 The key observation suggesting suppressed production of $q_i\bar{q}_j \,q\bar{q}$ exotics relative to
 the corresponding non-exotic $q_i\bar{q}_j$ is the following:
 The production of each $q\bar{q}$ pair during the first stage of the $M$ cluster production
 is suppressed. This suppression stems from the $\alpha_{(QCD)} \sim 1/N_c$ factor and is even stronger
 in nonperturbative  models e.g\cite{CNN} where the pairs are produced via Schwinger's mechanism.
 The $q\bar{q}q\bar{q}$ exotics are heavier than $\sim$ 1 GeV and should therefore be produced in this
 first stage when color singlet clusters of such masses are generated.
 Hence the production of exotics is suppressed relative to that of non-exotics by $1/N_c$ and most likely
 by much more.
 Similar reasoning  apply to decays of heavy quarks, i.e. B decays, despite the fact that the primary decay
 of the b quark to c and two light quarks yields three energetic quarks and we have also the spectator light
 quark from the initial B.

  As the final state pions are mainly "secondary" emerging from the decaying resonances/clusters, the relevant
 number of primary pairs initially produced is much smaller than the above naive estimate of ten pions and ten
 quark--anti-quark pairs per collision.

 We proceed next to discuss several cases where this pattern of suppressed production of exotics in electron
 positron colliders may manifest, and where it may help identify the state in question.

 i)The production of the new $X(3870)$ Belle state \cite {Belle} confirms this general pattern {\it if}
 it is exotic, say $c\bar{c}\, (u\bar{u} + d\bar{d})/2^{(1/2)}$.
 The original Belle experiment\cite{Belle} shows on the same $ (J/\psi) \pi\ pi$ invariant mass plot the
 (radially excited) non-exotic $^3\ S \; c\bar{c}\; \psi'$ state,and of $X(3870)$. Both states are expected
 to have branching fractions into $(J/\psi) \pi \pi$ of the same order of magnitude\cite{SW}. The large
 $\sim$ 300(!) ratio of the non-exotic and exotic peaks manifests then mainly the larger cross for producing
 the former.
  The interpretation of an exotic $X(3870)$ as a near threshold $D^*\bar{D}$ state \cite{Braaten2} and the
 Deuson model suggested for many other exotics\cite{Thornqist}  also imply strongly
 suppressed production.\cite{BKN},\cite{Gelman}

 ii) The BaBar $D_s (2317)$ state is strongly produced in B decay \cite{BaBar}. Thus our general considerations suggest
  that it is indeed the missing non-exotic  $c\bar{s}$ \cite{CJ}, rather than a $c\bar{s} q\bar{q}$
  \cite{BCL}.

 iii) The $ 0^+  \; a(980)$ and $f(980)$ states could be four quark states---as suggested
 by R.Jaffe\cite{Jaffe} or P-wave $0^{(++)} \; s\bar{s}$ non-exotics. In the first case these states should be less prominent
 in BaBar and Belle than $\omega,\rho$ or $ q{\bar q}$ P wave non-exotic resonances.

  iv) The new $D_s(2632)$ Selex state \cite{Selex} was not been seen to date in electron positron colliders.
 If it were exotic: $c\bar{s}(u\bar{u}+d\bar{d})/2^{(1/2)}$ then the suppressed production of exotics in
 $e^+e^-$ colliders provides (some) excuse for that.
 Other aspects of the Selex data argue more strongly against a non exotic $c\bar{s}$ assignment.
 The observed ratio $r=\Gamma(D^0K^+)/\Gamma(D_s^+\eta)$ is $ 0.16 \pm 0.06 $.
  However phase-space prefers the first higher Q value ,mode by $\sim$ factor of two, $s\bar{s}$
 production which must occur in the decay into $D_s + \eta$ ,is suppressed relative to the production of
 $u\bar{u} $ (or $d\bar{d}$) in the decay into $DK$ by  $\sim$ 3 and finally the $\eta$ is only $\sim$ 50\%
 $s\bar{s}$ in its flavor content. Jointly these three factors yield a predicted  $r$ (for a non
 -exotic $D^+_{sJ}(2632)) \sim 12$, seventy times larger than the observed value.

  The remaining puzzle of why this state with a large $Q$ value is so narrow is shared by the pentaquark,
 the prime exotic candidate to which we turn next.

  \section{Production of Exotics and Non-Exotics Off Nucleons}

 The non-production of complex hadronic structures, eg the He$^5$  nucleus and its anti-particle,
 in $e^+ e^-$ colliders is not an argument against their existence.
 One should look for He$^5$ in the natural neutron + He formation channel.
 A similar though weaker case is next made against using the lack of evidence for pentaquark in $e^+ e^-$ colliders
 as a reason to doubt its existence.

 (The lack of evidence for $J^P= 1/2^+ $ $\Theta$ and any of its expected  entourage of $1/2^-$,  $3/2^+$
 or other states in the natural $K^+$-neutron (namely $K^+$-deuteron scattering) formation channel, is however
 problematic \cite{N2}).

  It is well known and readily explained \cite{CNN} that the production of baryons in $e^+ e^-$ colliders is
 suppressed by a factor of~ 20 or more relative to that of mesons. This is much more the case for pentaquark
 production requiring five(!) pairs be produced.

 Within a specific model we found a dramatic realization of these general expectations.\cite{N2}
 (anti)Theta production in $e^+ e^-$ collisions was suppressed in this model by a large $10^5-10^6$ factor which may
 explain why the pentaquark has not seen to date in BaBar and in Belle.

   In $\gamma$-nucleon collisions, since at least three  (and if we
 allow the gamma to $s{\bar s}$ conversion even four) of the quarks required in order to make up the pentaquark are there initially, ${\Theta}$
 photoproduction should have far larger cross sections.
 The same is true in $N{\bar N}$ annihilations where we have altogether six initial quarks and antiquarks.
 In particular -in such  annihilations at rest or low energies we would expect to have often a diquark from the proton
 and an anti- diquark from the anti-nucleon to often form cryptoexotic $q q {\bar q} {\bar q} $ tetraquarks. If
 the latter were reasonably narrow than the tetraquarks should have been discovered in the famous CERN LEAR experiment.
 If however the lightest (and narrowest!) tetraquarks are significantly lighter than W(annihilation)$ {\approx 2}$ GeV
 several pions
 accompany on average the tetraquark in each annihilation event generate a severe
 combinatorial background impeding such a discovery\cite{GN},\cite{CN}.

 The above comments notwithstanding, many features of the pentaquark and its
 production pattern remain puzzling.

 The above strong suppression obtained by using the chromoelectric flux tube model (CFT) for particle production
 in electron positron collisions\cite{CNN} {\it and} a CFT model with two junctions and one anti-junction for
 the pentaquark.\cite{CN},\cite{GN}
 This CFT model for the pentaquark was motivated by the small width  $\Gamma \sim$ O(MeV) suggesting that
 the pentaquark is very different from the decay channel hadrons. The model naturally corresponds to
 the diquark-diquark--anti-strange-quark picture for the pentaquark (\cite{JW} and also \cite{N1}).
 This picture along with some extensions\cite{N2} of QCD inequalities \cite{NL} imply new undiscovered vector
 meson tetraquarks lighter than a GeV. Furthermore in the CNN CFT model the pentaquark in $\gamma$ nucleon
 collisions should be associated with such tetraquarks and {\it not} with the observed Kaons.

 Also independently of any specific model we have argued \cite{CN} that the pentaquark production in the various
 photoproduction reactions of the pentaquark are inconsistent with the natural kaon exchange models: the large
 $\Theta-KN$ couplings required clash with  bounds on the width of the
 pentaquark \cite{N1},\cite{ASW},\cite{HC},\cite{CT}.

  In general, meson baryon is the ideal formation channels of pentaquarks as the initial state contains the required
 four quarks and antiquark.
 In $KN$ collisions no non-exotic $s$ channel states exist and even in ${\bar K}N$ or $\pi N$ collisions the total
 crossection is dominated for most energies by exotic namely pentaquark intermediate states.
 Conventially these state were assumed to be very broad. Further the exotic spectrum was expected to be denser
 than the non-exotic reflecting the larger number of degrees of freedom. The many overlapping ressonances would then blend
 into the smooth energy variation observed. If the pentaquark survives this view may be challanged: many ressonances
 may have been too narrow and missed in past rough scans \cite{GN}. Also if some pentaquarks are indeed three
 body diquark diquark anti-quark systems their spectrum need not be (much) denser than that of baryons.

 One final comment on the use of nuclear targets. It is well known that the charge radius of Nucleons is larger than that
 of mesons and multi-quark exotics are likly to be even more extended. Thus  heavy nuclear targets may
 filter out---particularly at higher collision energies, the exotics and more readily pass smaller more compact mesons.
 Under special circumstances this may be evaded. In the CFT model some tetra-penta etc quarks are topologically stable
 and will not break while traversing the nucleus.
 Also for low collision energies the many initial quarks can even enhance production of multi-quark structures.

  \section{Some Comments on the Selex Doubly-Charmed Baryons and on QCD Inequalities}

  The doubly-charmed baryons provide yet another example of conflicting experimental evidence for new states.
 Can theory provide some additional hints?
 Earlier higher theoretical estimates for the mass of the doubly charmed baryons cannot exclude the Selex
 discovery. Indeed prior to the discovery of the $D_s(2317)$ BaBar  $0^{(++)}$ state most theoretical estimates of
 its mass were $\sim$ 50-100 MeV higher (and its expected broad $K-D$ width discouraged searching for it...)

 However the large EM mass splitting in the Doubly-Charmed Baryon (DCB) doublet:
\[
\delta(m)| DCB= m_{(ccu)} -m_{(ccd)} \sim 12 {\rm  MeV}
\]
is problematic. It is inconsistent with the splitting in the Charmed Meson (CM) doublet:
\[
\delta(m)| CM = m_{(c\bar{d})} - m_{(c\bar{u})} = 4.8 \pm 0.1
{\rm MeV}
\]
 Detailed QCD/potentials modeling and quarkonium phenomenology \cite {RQ} imply an avarage $c-{\bar c}$
 separation in$ J/\psi$ of ~.4 Fermi. The cc-diquark system inside the  DCB is bound by {\it half} the
 color forces operative in the $J/\psi$, and hence the cc system is larger than $J/\psi$.

 An extreme assumption helping compare the EM mass splittings in the DCB and in the CM is that the
 cc diquark is a pointlike ``Heavy'' system of twice the mass and charge of the charmed quark.
 {This grossly distorts the cc EM self energy---which, however, cancels in the ccu-ccd difference). Concentrating
 the heavy quark charges at one point and using the same, universal, wave function of the light anti-quark
 or quark for heavy color source mass $m_c$ or $2m_c$, both enhance the EM splitting.
 Since  we look for an {\it upper } bound on this splitting we adopt both approximations.
 The electromagnetic part of the isodoublet splitting in the DCB is therefore at most twice that in
 the CM. Using $m(d)-m(u) \sim$  3-5 MeV we predict that $\delta(m)$ |DCB $< 3.6-0$ MeV conflicting with Selex data.

  Before concluding we recall some simple, semi-empirical rules that evolved from from QCD inequalities due to
 Weingarten,\cite{W} Vafa and Witten \cite{VW} and Witten\cite {Witten} and Nussinov \cite{N3}
 This subject was extensively reviewed in a recent report \cite {NL}.

 Let us mention here a few pertinent examples. The pseudoscalar mass inequalities
\[
 m_{ps}(q_i{\bar q}_j) >1/2[m_{ps}(q_i {\bar q}_i)+ m_{ps}(q_j {\bar q}_j)]
\]
 follow directly from the QCD lagrangian if we neglect ``Flavor disconnected''
 diagrams with intermediate pure glue states. This is indeed justified for$ q= Q= c$ or b . Hence the lightest $ B_c 0^-$ state with reported mass of 6.4+-.4 GeV,
 \cite {PDG} {\it must} be heavier than the avarage of the $ \eta_c$ and $\eta_b$ masses. Hopefully the $\eta_b$ will
 be discovered soon and this QCD prediction verified.
 More heuristic analoge inequalities are expected for the $3^S$ vector mesons made of heavy quarks.
 suggesting a symilar inequality involving $ \Upsilon ^4 S, J/\psi$ and the lightest vector $B_c$ state.
  Doubly-charmed baryons should satisfy the meson baryon inequalities:$ m (ccq_i)> 1/2[m(J/\psi)+ a.m (D^*_i)+ (2-a).m (D_i ]$ with a=2 for S= 3/2 and a=1/2 for S=1/2  DCB's.(S=spin) \cite {NL}
  In the baryon sector various convexity relations between baryon masses have been motivated.
 For multiple charmed baryons these imply: $ m(cud)=m(\Lambda_c)>1/2[mccq+m{\rm(Nucleon)}]$  and
 $ m(ccu)(3/2^+) > (1/2)m(ccc)(3/2)^+ + m(cud)(3/2)^+$
  Jointly thses relations bracket the mass of the lightest $S=1/2$ DCB in the range of 3 GeV-3.63 GeV.

  \section{Summary}

  Despite recent progress in Lattice QCD we still lack reliable ab initio calculations of hadron masses
 particularly with one or more light ($u,d,s$) quarks. In looking for guidance as to where new
 discoveries are likely and/or for ``theoretical confirmation'' of putative findings experimenters often turn to
 the many hadronic models developed prior to and alogside QCD. These include
 potentia models for massive constituent quarks, chiral lagrangians and chiral pertrubations, and QCD sum rules.
  Some intriguing offshoots of the chiral approach and large $N_c$ limit are the Skyrme/chiral soliton models.
 In certain extensions of the original SU(2) flavor model to SU(3) a very light \cite{Pr} and narrow \cite {DPP}
 $1/2 ^+$ pentaquark state emerges as part of an SU(3) antidecuplet.
 The last work, which has been subsequently challenged \cite {Cohen} and \cite {Itzhaki}, motivated Nakano et al.
 to embark on their discovery work.
 A different example is provided by the $D_s$ states. After discovery it was realized that such states were suggested
 in a special spontaneous chiral symmetry breaking pattern were some parity doublet regulatities survive \cite{NZBEH}

 In the present and in many earlier works we note that more basic/elementary considerations of unitarity, heavy quark
 universality, QCD inequalities and general semi-empirical patterns may be critical in assesing discoveries of new
 hadronic states.
 Also a conflicting patterns of production of exotic and non-exotic resonances in various reactions need not
 imply that some of the experiments are wrong and may instead be due to rather simple underlying physics.

 \section{Acknowledgements}

  I would like to thank M. Purohit for bringing to my attention the new $D_s$ Selex state,
  and to him and to C.Rosenfeld
 and J. Wilson for careful reading and useful comments on the manuscript.

\end{document}